\documentclass[10pt,aps,prd,showpacs,one-column,nofootinbib,noshowkeys,floatfix,preprintnumbers]{revtex4-2}
\usepackage{graphics,graphicx}
\usepackage{amsmath, amssymb}
\usepackage{multirow}
\usepackage{color}
\usepackage{verbatim}
\usepackage{hyperref}
\usepackage[normalem]{ulem}  
\usepackage{ulem}
\usepackage{color}
\usepackage{cancel}
\usepackage{mathtools}
\usepackage{epstopdf}
\usepackage{url}
\usepackage{footnote}
\usepackage[absolute]{textpos}

\newcommand{\nbcan}{\langle N_B\rangle}
\newcommand{\nabcan}{\langle N_{\bar{B}}\rangle}

\begin{document}


\title{Fluctuations in the canonical ensemble of an Abelian charge}
\author{Bengt Friman}
\affiliation{GSI Helmholtzzentrum f{\"u}r Schwerionenforschung, 64291 Darmstadt, Germany}
\author{Krzysztof Redlich}
\affiliation{Institute of Theoretical Physics, University of Wroc\l{}aw,  50204 Wroc\l{}aw, Poland}
%



\begin{abstract}
  We study fluctuations in the canonical ensemble, where the net baryon number is exactly conserved. The focus is on cumulants and factorial cumulants linked to the  baryon and antibaryon multiplicities and their sum or difference in full phase-space as well as in subsystems.  In particular, we connect the fluctuations of the net baryon number in a subsystem, relevant for fluctuation studies in nucleus-nucleus collisions,  with fluctuations of the baryon and antibaryon numbers of the total system. We derive analytic expressions for factorial cumulants of arbitrary order.  Compact results are obtained in terms of cumulants of the baryon number of the total system. Moreover, we derive the  asymptotic forms of the factorial cumulants of baryon and antibaryon multiplicities in the high- and low-temperature limits and discuss the results in the context of heavy-ion collision experiments.
\end{abstract}
\maketitle 

\section{Introduction}
One of the goals of current experimental and theoretical studies of chiral symmetry restoration in QCD is to unravel the phase diagram of strongly interacting matter and to verify whether a chiral critical endpoint exists. A dedicated research program at RHIC, the beam energy scan (BES), has been established to explore these issues in collisions of heavy ions at relativistic energies \cite{Aggarwal:2010cw}. 
Fluctuations and correlations  of conserved  charges are considered as possible probes of the QCD phase diagram~
\cite{Stephanov:1998dy,Stephanov:1999zu,Asakawa:1989bq,Friman:2011pf,Ejiri:2004bh,Allton:2005gk,Ejiri:2005wq,Karsch:2005ps,Karsch:2010ck,Sasaki:2007db,Sasaki:2006ww,Bzdak:2019pkr,Kuznietsov:2022pcn}.
These  are experimentally accessible  observables and reflect the criticality of the chiral transition. Fluctuations of the net baryon number are particularly interesting, owing to a direct connection to critical scaling near the chiral phase boundary~\cite{Friman:2011pf,Karsch:2010ck}. Other useful quantities characterizing multiparticle-correlations, and possibly also the chiral phase transition, are factorial cumulants of the baryon and antibaryon multiplicity distributions \cite{Bzdak:2019pkr,Barej:2020ymr,Bialas:2007ed,Bzdak:2018axe,Barej:2022jij}.
Data on proton-, antiproton- and net-proton-number fluctuations  in heavy-ion collisions have been obtained by the STAR Collaboration at  RHIC ~\cite{Adamczyk:2013dal,Luo:2015ewa,Luo:2015doi,STAR:2021iop} and the ALICE Collaboration~ \cite{Rustamov:2017lio,Arslandok:2020mda,Acharya:2019izy} at  LHC energies. These are utilized as proxies for fluctuations of the baryon, antibaryon and net baryon numbers, respectively. The corresponding results on the variance, skewness, and kurtosis of the net proton number are intriguing and have triggered discussions on their physics origin and interpretation, focused mainly on the connection to chiral criticality. 

However, an important aspect of the analysis of such fluctuations 
in heavy-ion collisions, is that the resulting cumulants are affected also by fluctuations unrelated to criticality.  Two such effects are of particular relevance:   volume fluctuations~\cite{Skokov:2012ds,Braun-Munzinger:2016yjz,Sugiura:2019toh}, which are linked to event-by-event fluctuations of the number of participating nucleons,  and constraints  imposed  on fluctuation observables by exact  conservation of the net baryon number in full phase space 
\cite{Bzdak:2019pkr,Barej:2020ymr,Bialas:2007ed,Bzdak:2018axe,Barej:2022jij,Braun-Munzinger:2016yjz,Bzdak:2012an,Braun-Munzinger:2018yru,Braun-Munzinger:2019yxj,Begun:2004gs,Braun-Munzinger:2020jbk,Vovchenko:2020tsr, Vovchenko:2020gne,Pruneau:2019baa}

Recently, the constraints on fluctuations set by the global conservation of additive charges, in particular of the net baryon number, were explored for current experimental setups.  In these studies cumulants of net-baryon-number fluctuations
\cite{Bzdak:2019pkr,Braun-Munzinger:2016yjz,Bzdak:2012an,Braun-Munzinger:2018yru,Braun-Munzinger:2019yxj,Braun-Munzinger:2020jbk,Vovchenko:2020tsr,Vovchenko:2020gne}, as well as factorial cumulants of the baryon and antibaryon multiplicity distributions
\cite{Bzdak:2019pkr,Barej:2020ymr,Barej:2022jij} were considered. It was shown that  the magnitude and energy dependence of the suppression of net-proton number cumulants relative to the Skellam baseline, observed by the  STAR and ALICE collaborations in ultra-relativistic nucleus-nucleus collisions, is, given the present experimental uncertainties, consistent with exact conservation of the total net baryon number  \cite{Braun-Munzinger:2019yxj, Braun-Munzinger:2020jbk}.

In the following, we extend the so far published results on  baryon conservation effects on fluctuation observables. In particular, we connect the fluctuations of the net baryon number in a subsystem, relevant for fluctuation studies in nucleus-nucleus collisions, with fluctuations of the baryon and antibaryon numbers in the total system. Furthermore, taking into account exact conservation of the baryon number, we compute factorial cumulants of the baryon, antibaryon,  mixed baryon-antibaryon and net baryon number multiplicity distributions  in a subsystem.  We derive general expressions for these factorial cumulants up to arbitrary order.  Compact results are obtained in terms of the cumulants of the baryon number of the total system.  These results are an extension of the previous finding by Barej and Bzdak~\cite{ Barej:2020ymr,Barej:2022jij}, where analytic results for the corresponding factorial cumulants up to sixth order were derived.  

Considering possible applications of our results in  heavy-ion collisions  we also derive  asymptotic forms for the factorial cumulants of baryon and antibaryon multiplicities in two limits, which correspond to the conditions in heavy-ion collision experiments at high and low beam energies, respectively.  The analytic expressions for cumulants and factorial cumulants of arbitrary order, as well as their relations, allow a systematic approach to  fluctuations in the canonical ensemble, which in turn provide an important baseline for the analysis of fluctuation data in heavy-ion collisions.    

The paper is organized as follows: In the next Section we introduce the canonical partition function for uncorrelated baryons, where the net baryon number is conserved  and  discuss pertinent fluctuation observables. In Section 3 we formulate the cumulant and factorial cumulant generating functions. Analytic  expressions for factorial cumulants of the baryon,   antibaryon, and  the net baryon numbers in a subsystem  are obtained in Section 4, while the high- and low-energy limits of the cumulants and factorial cumulants are presented in Section 5. We summarize our results in  Section 6. Mathematical details and general results, which are independent of the assumed partition function, are presented in three appendices.


\section{Canonical ensemble of  the net baryon number}
To compute the influence of exact charge conservation on fluctuation observables, we adopt a thermal model for the net baryon number conservation in the Boltzmann approximation and neglect baryon-baryon interactions.  The statistical operator is formulated following 
the S-matrix approach \cite{Dashen:1969ep,Venugopalan:1992hy,Weinhold:1997ig,Giacosa:2016rjk,Lo:2017sde,Dash:2018mep,Friman:2015zua, Lo:2020phg}, where to leading order in the fugacity expansion it has the form of an ideal gas, albeit with  the thermal phase-space of free baryons modified by meson-baryon interactions. 

The S-matrix thermodynamic potential 
reproduces particle production yields in heavy-ion collisions  \cite{ Braun-Munzinger:2003pwq,Andronic:2017pug,Cleymans:2020fsc, Andronic:2018qqt}, and 
describes basic properties of the net proton number cumulants and their energy dependence, as obtained by the STAR Collaboration \cite{ Braun-Munzinger:2020jbk}.  Moreover, the S-matrix thermodynamic potential of a hadron gas formulated in the grand canonical ensemble is consistent with the lattice QCD  equation of state and some second-order cumulants and correlations of conserved charges in the confined phase \cite{Bazavov:2017dus, Noronha-Hostler:2019ayj, Lo:2017lym,Almasi:2019yaw,Goswami:2020yez}.

In the grand canonical partition function
\begin{eqnarray}\label{eq:grand-canonical-partition}
    \mathcal{Z}(\mu_B,T)&=&\sum_{N_{B}=0}^{\infty}\sum_{N_{\bar{B}}=0}^{\infty}\frac{(e^{\mu_B/T}\,z_B)^{N_{B}}}{N_{B}!}\frac{(e^{-\mu_B/T}\,z_{\bar{B}})^{N_{\bar{B}}}}{N_{\bar{B}}!} \nonumber\\
    &=&\exp\big(e^{\mu_B/T}\,z_B+e^{-\mu_B/T}\,z_{\bar B}\big).
\end{eqnarray}
the fluctuations of baryons and antibaryons are described by two independent Poisson distributions. 
In the spirit  of the S-matrix approach, the effect of meson-baryon interactions is subsumed in the baryon and antibaryon single-particle partition functions $z_B$ and $z_{\bar B}$. 

The canonical partition function in the ensemble, where the net baryon number is conserved, is given  by \cite{Braun-Munzinger:2020jbk,Braun-Munzinger:2003pwq} \footnote{As noted in~\cite{Braun-Munzinger:2020jbk}, this form of the partition function applies also to systems where the composition is non-uniform, provided the baryon and antibaryon multiplicities are locally Poisson distributed. Consequently, it is more general than the global thermal statistical model employed as a motivation.}
\begin{eqnarray}
    Z_{B}&=&\sum_{N_{B}=0}^{\infty}\sum_{N_{\bar{B}}=0}^{\infty}\frac{(\lambda_{B}\,z_B)^{N_{B}}}{N_{B}!}\frac{(\lambda_{\bar{B}}\,z_{\bar{B}})^{N_{\bar{B}}}}{N_{\bar{B}}!}\delta({N_{B}-N_{\bar{B}}-B}) \nonumber\\
    &=&\int_0^{2\pi}\frac{d\phi}{2\pi}\,e^{-iB\phi}\exp\left({\lambda_B\,z_B\,e^{i\phi}+\lambda_{\bar B}\,z_{\bar{B}}\,e^{-i\phi}}\right)
    \label{eq:canonical-partition}\\
    &=&\left(\frac{\lambda_{B}\,z_B}{\lambda_{\bar{B}}\,z_{\bar{B}}}\right)^{\frac{B}{2}}\,I_{B}(2\,z\,\sqrt{\lambda_{B}\,\lambda_{\bar{B}}}),\nonumber
\end{eqnarray}
where the auxiliary parameters  $\lambda_{B,\bar{B}}$ are introduced for the calculation of the mean number of baryons and antibaryons and the corresponding cumulants,
\begin{eqnarray}\label{NCM}
    \langle N_B\rangle&=& \lambda_B{\frac{\partial\ln Z_B} {\partial\lambda_B}}|_{\lambda_B,\lambda_{\bar B}=1}=z\,\frac{I_{B-1}(2\, z)}{I_{B}(2\, z)},\\\label{NCP}
    \langle N_{\bar{B}}\rangle&=&\lambda_{\bar B}{\frac{\partial\ln Z_B} {\partial\lambda_{\bar B}}}|_{\lambda_B,\lambda_{\bar B}=1}=z\,\frac{I_{B+1}(2\, z)}{I_{B}(2\, z)},\\
    \langle (\delta N_{\rm B})^k\rangle_c&=&\left[\left(\lambda_B\,\frac{\partial}{\partial  \lambda_B}\right)^{k}\log Z_B\right]_{\lambda_B,\lambda_{\bar B}=1}\qquad (k>1).\label{eq:cumulant-lambda-deriv}
\end{eqnarray}
Here (and below) $\langle \dots \rangle$ denotes an expectation value in the canonical ensemble, while $\langle \dots \rangle_c$ is the connected part thereof. Thus, fluctuations of the baryon and antibaryon numbers in the canonical ensemble are given by $\delta N_B=N_B-\langle N_B\rangle$ and $\delta N_{\bar B}=N_{\bar B}-\langle N_{\bar B}\rangle$,
and (\ref{eq:cumulant-lambda-deriv}) yields the cumulants of the baryon number in the full system.

We can rewrite (\ref{eq:cumulant-lambda-deriv}) in terms of the mean baryon and antibaryon mulitpicities, keeping the dependence on the auxiliary parameters $\lambda_{\rm B},\lambda_{\bar{\rm B}}$
\begin{equation}\label{eq:cumulant-recurrence}
    \langle (\delta N_{\rm B})^k\rangle_c=\frac12\,\left[\left(\lambda_B\,\frac{\partial}{\partial  \lambda_B}\right)^{k-1}\left(\langle N_B\rangle_\lambda+\langle N_{\bar{B}}\rangle_\lambda\right)\right]_{\lambda_B,\lambda_{\bar B}=1}\quad (k>1),
\end{equation}
where
\begin{eqnarray}\label{NCM-2}
    \langle N_B\rangle_\lambda&=&z\,\sqrt{\lambda_B\,\lambda_{\bar B}}\,\frac{I_{B-1}(2\, z\,\sqrt{\lambda_B\,\lambda_{\bar B}})}{I_{B}(2\, z\,\sqrt{\lambda_B\,\lambda_{\bar B}})},\\\label{NCP-2}
    \langle N_{\bar{B}}\rangle_\lambda&=&z\,\sqrt{\lambda_B\,\lambda_{\bar B}}\,\frac{I_{B+1}(2\, z\,\sqrt{\lambda_B\,\lambda_{\bar B}})}{I_{B}(2\, z\,\sqrt{\lambda_B\,\lambda_{\bar B}})}.
\end{eqnarray}
Hence, the baryon and antibaryon multiplicities, (\ref{NCM-2}) and (\ref{NCP-2}), are functions of $z\sqrt{\lambda_B\,\lambda_{\bar B}}$. This reflects the fact that in the canonical ensemble $N_B-N_{\bar B}$ is conserved and thus does not fluctuate~\footnote{This is easily seen by noting that the recurrence relation $2\,\nu\, I_\nu(x)=x\,\big[I_{\nu-1}(x)-I_{\nu+1}(x)\big]$ implies that $\langle N_B\rangle_\lambda-\langle N_{\bar{B}}\rangle_\lambda=B$ is independent of $\lambda_B$ and $\lambda_{\bar B}$.}. It follows that the fluctuations of $N_B$ and $N_{\bar B}$ are equal, i.e., $\delta N_{\rm B}=\delta N_{\bar{\rm B}}$ and that cumulants of the form
\begin{equation}\label{eq:canonical-cumulant-eq}
    \langle (\delta N_{\rm B})^{n-m}\,(\delta N_{\bar{\rm B}})^m\rangle_c
\end{equation}
are independent of $m$, and thus all equal to $\langle (\delta N_{\rm B})^{n}\rangle_c$. In the {\em full} system, it is therefore sufficient to consider only cumulants of the baryon number. 

Moreover, it follows that the derivatives of $\langle N_B\rangle_\lambda+\langle N_{\bar{B}}\rangle_\lambda$ with respect to $\lambda_B$ can be replaced by derivatives with respect to $z$, setting $\lambda_B=\lambda_{\bar B}=1$. 
Using the notation
\begin{align}\label{eq:shorthand-1}
    c_1=&\frac12 \langle N_{\rm B} + N_{\bar{\rm B}}\rangle,\\
    c_k=&\langle (\delta N_{\rm B})^k\rangle_c\qquad (k>1), \label{eq:shorthand-2}
\end{align}
we can then rewrite (\ref{eq:cumulant-recurrence}) as
\begin{equation}\label{eq:cumulants-NB}
    c_k=\left(\frac{z}{2}\,\frac{d}{d  z}\right)^{k-1}c_1\qquad (k>1).
\end{equation}
Clearly, the cumulants satisfy the recurrence relation
\begin{equation}\label{eq:rec-rel-n}
    c_{k+1}=\frac{z}{2}\,\frac{d}{d z}\, c_k.
\end{equation}

It is useful to define $N^{(tot)}=N_B+N_{\bar B}$ and to consider the  fluctuations thereof, $\delta N^{(tot)}=\delta N_B + \delta N_{\bar B}$. We introduce the notation
\begin{eqnarray}\label{eq:cumulants-Ntot}
    C_1&=&\langle N^{(tot)}\rangle,\\
    C_k&=&\langle (\delta N^{(tot)})^k\rangle_c \qquad (k>1). \nonumber
\end{eqnarray}
Using the fact that $\delta N_{\rm B}=\delta N_{\bar{\rm B}}$ and (\ref{eq:shorthand-1},\ref{eq:shorthand-2},\ref{eq:rec-rel-n}), it follows that
\begin{equation} \label{eq:relation-cumulants}
    C_k=2^k\,c_k
\end{equation}
 and
\begin{equation}\label{eq:rec-rel-N}
    C_{k+1}=z\,\frac{d}{d z}\, C_k.
\end{equation}

It is straighforward to compute the baryon number cumulants $c_k$ using  the recurrence relation for $c_k$, (\ref{eq:rec-rel-n}). Given that $c_1=\frac12(N_B+N_{\bar B})$, and using the notation~\cite{Braun-Munzinger:2020jbk}
\begin{eqnarray}\label{eq:SPQW}
    S&=&\langle N_B+N_{\bar B}\rangle,\nonumber\\
    P&=&\langle N_B\rangle\,\langle N_{\bar B}\rangle,\\
    Q&=&z^2-P,\nonumber\\
    W&=&Q\,S-P.\nonumber
\end{eqnarray}
with the derivatives
\begin{equation}\label{eq:SPQW-deriv}
    S'=\frac{4}{z}\,Q,\qquad  P'=\frac{2}{z}\,Q\,S,\qquad Q'=\frac{2}{z}\,\big(Q-W\big),
\end{equation}
one finds for the first few
\begin{eqnarray}\label{eq:n_S_Q-W}
    c_1&=&S/2,\nonumber\\
    c_2&=&Q,\\
    c_3&=&Q-W,\nonumber\\
    c_4&=&Q-W\,+W\,S-2Q^2.\nonumber
\end{eqnarray}
The corresponding expressions for the cumulants $C_k$ are trivially obtained using (\ref{eq:relation-cumulants}).

\section{Cumulant generating functions}\label{sect:cum-gen-functs}

The generating function for the cumulants of $N^{(tot)}=N_B+N_{\bar B}$, (\ref{eq:cumulants-Ntot}),  in the canonical ensemble with net baryon number $B$ is obtained by evaluating
\begin{eqnarray}\label{eq:can_cumulant_gen_function}
    G_c(t)&=&\ln\Big[\sum_{N_{B}=0}^{\infty}\sum_{N_{\bar{B}}=0}^{\infty}\frac{(z_B)^{N_{B}}}{N_{B}!}\frac{(z_{\bar{B}})^{N_{\bar{B}}}}{N_{\bar{B}}!}\delta({N_{B}-N_{\bar{B}}-B})\,e^{(N_B+N_{\bar B})\,t}\Big]\nonumber\\
    &=&\ln\Big[\int_0^{2 \pi}\frac{d \phi}{2\,\pi}\,\exp\big(z_B\,e^{t+i \phi}\big)\,\exp\big(z_{\bar B}\,e^{t-i \phi}\big)\,e^{-i\,\phi\,B}\Big]\\
    &=&\ln\Big[\left(\frac{z_B}{z_{\bar B}}\right)^{B/2}\,I_B\big(2\,z\,e^t\big)\Big],\nonumber
\end{eqnarray}
where we set the auxiliary parameters $\lambda_B,\lambda_{\bar B}$ to unity. The $k$:th cumulant is then given by
\begin{equation}\label{eq:can-cumulant-Ntot}
    C_k=\frac{d^k G_c(t)}{d\,t^k}\mid_{t=0}.
\end{equation}

The generating functions for the corresponding factorial cumulants is obtained by the replacement~\cite{Kitazawa:2017ljq} $e^t\to x$,
\begin{equation}
    H_c(x)=\ln\Big[\left(\frac{z_B}{z_{\bar B}}\right)^{B/2}\,I_B\big(2\,z\,x\big)\Big].
\end{equation}
The factorial cumulants of $\delta(N_B+N_{\bar B})$ are then given by
\begin{equation}\label{eq:factorial-cumulants-tot}
    F_n=\frac{d^n H_c(x)}{d\,x^n}\mid_{x=1}.
\end{equation}
We note that the factorial cumulants are identical to the functions $f^{(n)}(z)$ defined in \cite{Braun-Munzinger:2020jbk} and thus satisfy the recurrence relation
\begin{equation}\label{eq:rec-relation-F}
F_{n+1}=z^{n+1}\frac{d}{d\,z}\left(F_{n}/z^n\right).
\end{equation}
The factorial cumulant $F_n$ of any order $n$ can be obtained by using (\ref{eq:rec-relation-F}) and (\ref{eq:SPQW},\ref{eq:SPQW-deriv}), starting from $F_1=C_1=S$.

The cumulants and factorial cumulants in the full system are connected via the general relations \begin{equation}\label{eq:Cn_to_Nk-relation}
    F_n=\sum_{k=1}^k\,s(n,k)\,C_k,
\end{equation}
and
\begin{equation}\label{eq:Nn-Ck-relation}
     C_n=\sum_{k=1}^n\,S(n,k)\,F_k,
\end{equation}
derived in~\ref{sect:App-c}, where $s(n,k)$ and $S(n,k)$ are Stirling numbers of the first and second kind, respectively. 

\section{Fluctuations in a subsystem}

In this section we explore the fluctuations in a subsystem, $A$, of the canonical system discussed so far. The part of the system that is not in $A$ we denote by $R$. While the net baryon number is conserved in the full system, $A+R$, it is not in the subsystem, because net baryon number can be transferred between $A$ and $R$. We obtain analytic results for the cumulants and factorial cumulants of the baryon and antibaryon numbers in $A$. Moreover, we derive general relations between these cumulants and those of the net baryon number. 

The cumulants of fluctuations in a subsystem are computed by differentiating the corresponding generating functions. Thereby we employ the Fa\`{a} di Bruno formula~\cite{Riordan:1946,Comtet:1974}
\begin{equation}\label{eq:faa-di-bruno-1}
    \frac{d^n}{dx^n}f(h(x))=\sum_{k=1}^n f^{(k)}(h(x))\,B_{n,k}\left(h^{(1)}(x),h^{(2)}(x),\dots,h^{(n-k+1)}(x)\right),
\end{equation}
and generalizations thereof~\cite{Riordan:1946,Schumann:2019xy}.
In (\ref{eq:faa-di-bruno-1}), $f^{(k)}$ and $h^{(k)}$ denote the $k$th derivatives while $B_{n,k}(y_1,y_2,\dots)$ are partial Bell polynomials~\cite{Bell:1927}.

\subsection{Factorial cumulants of \texorpdfstring{$\delta N_B$}{d N B} and \texorpdfstring{$\delta N_{\bar B}$}{d N B}}\label{sec:fac-cum}

Analytic expressions for the factorial cumulants of the baryon and antibaryon numbers up to sixth order in a subsystem, taking the exact conservation of the net baryon number in the total system into account, were presented in~\cite{Barej:2020ymr}. In this section we derive closed-form expressions for these factorial cumulants. Compact expressions are obtained in  terms of the cumulants of the baryon number of the total system, $c_k$.

The generating function for factorial cumulants is in this case~\cite{Barej:2020ymr}
\begin{eqnarray}\label{eq:generating-function-fact-cum}
    g_{fc}(x,\bar{x})&=&\frac{B}{2}\big[\ln\big(p(x)\big)-\ln\big(\bar{p}(\bar{x})\big)\big]\\
    &+&\ln\big[I_B\big(2\,z\,\sqrt{p(x)\,\bar{p}(\bar{x})}\big)\big],\nonumber
\end{eqnarray}
where $B$ is the total net baryon number, $p(x)=1-\alpha_B+\alpha_B\,x$ and $\bar{p}(\bar{x})=1-\alpha_{\bar B}+\alpha_{\bar B}\,\bar{x}$. Here  $\alpha_B$ is the probability for finding a baryon in the subsystem $A$ and $\alpha_{\bar B}$ that for an antibaryon, while $z=\sqrt{z_B\,z_{\bar{B}}}$ is the geometric mean of the single-particle partition functions $z_B$ and $z_{\bar{B}}$ (cf. Ref.~\cite{Braun-Munzinger:2020jbk}).

The factorial cumulants in the subsystem $A$ are obtained by evaluating the derivatives with respect to $x$ and $\bar{x}$
\begin{equation}\label{eq:factorial-cumulants}
    F^{(n,m)}=\frac{\partial^n}{\partial\,x^n}\,\frac{\partial^m}{\partial \bar{x}^m}\,g_{fc}(x,\bar{x})|_{x=\bar{x}=1}.
\end{equation}

The derviatives of the first two terms in (\ref{eq:generating-function-fact-cum}), $g_{fc,1}(x)$ and $g_{fc,2}(\bar{x})$, are obtained by employing the Fa\`{a} di Bruno formula for functions of the form $\log(f(y))$, where $f(y)$ is a first order polynomial in y~\cite{Comtet:1974}. One finds
\begin{equation}
    \frac{\partial^n}{\partial x^n}\,g_{fc,1}(x)=-\frac{B}{2}\,(-\alpha_B)^n\,(n-1)!,
\end{equation}
and
\begin{equation}
    \frac{\partial^n}{\partial \bar{x}^n}\,g_{fc,2}(\bar{x})=\frac{B}{2}\,(-\alpha_{\bar B})^n\,(n-1)! .
\end{equation}

For the last term in (\ref{eq:generating-function-fact-cum}),
\begin{equation}\label{eq:fc-gen-func-c-term}
    g_{fc,3}(x,\bar{x})=\ln\big[I_B\big(2\,z\,\sqrt{p(x)\,\bar{p}(\bar{x})}\big)\big],
\end{equation}
we consider the composite function $f_3( g_3(x,\bar{x}))$, where $f_3( y)=\ln\big[I_B(2\,z\,e^y)\big]$ and $g_3(x,\bar{x})=\frac12\,\big(\ln\big[p(x)\big]+\ln\big[\bar{p}(\bar{x})\big]\big)$. The derivatives of $f_3(y)$ are obtained using (\ref{eq:can_cumulant_gen_function}) and (\ref{eq:can-cumulant-Ntot}), which yield $f^{(n)}_3(0)= C_n$. Moreover, the non-zero derivatives of $g_3(x,\bar{x})$ are given by
\begin{eqnarray}\label{eq:gc-derivatives}
    g_3^{(n,0)}&=&-\frac12(-\alpha_B)^n\,(n-1)!\quad (n>0)\\
    g_3^{(0,m)}&=&-\frac12(-\alpha_{\bar B})^m\,(m-1)!\quad (m>0).\nonumber
\end{eqnarray}
Now, using the multivariate Fa\`{a} di Bruno formula to compute the derivatives with respect to $x$ and $\bar{x}$ of $g_{fc,3}(x,\bar{x})$ we find
\begin{equation}
    \frac{\partial^n}{\partial\,x^n}\,\frac{\partial^m}{\partial \bar{x}^m}\,g_{fc,3}(x,\bar{x})=\sum_{k=1}^{n+m}\,C_k\,B_{n,m;k}(\{g_3^{(i,j)}\})
\end{equation}
where the $B_{n,m;k}(\{g_3^{(i,j)}\})$ are multivariate Bell polynomials (see  \ref{sec:app-a} and Ref.~\cite{Schumann:2019xy}). As shown in \ref{sec:app-b}, the multivariate Bell polynomials are, for the derivatives (\ref{eq:gc-derivatives}), given by
\begin{equation}
    B_{n,m;k}(\{g_3^{(i,j)}\})=\frac{(\alpha_B)^n\,(\alpha_{\bar B})^m}{2^k}\,\,s(n,m;k),
\end{equation}
where $s(n,m;k)$ are generalized Stirling numbers of the first kind. They 
can be expressed in terms of the standard Stirling numbers of the first kind,
\begin{equation}
    s(n,m;k)=\sum_{l=0}^{k}\,s(n,k-l)\,s(m,l).
\end{equation}

Now, collecting all terms and using (\ref{eq:relation-cumulants}), we find a closed-form expression for the factorial cumulants of a subsystem, which accounts for baryon-number conservation in full phase space,
\begin{eqnarray}\label{eq:factorial-cumulants-closed-form}
    F^{(n,m)}&=&-\frac{B}{2}\Big((-\alpha_B)^{n}\,(n-1)!\,\delta_{m,0}-(-\alpha_{\bar B})^{m}\,(m-1)!\,\delta_{n,0}\Big)\nonumber\\
    &+&(\alpha_B)^n\,(\alpha_{\bar B})^m\sum_{k=1}^{n+m}c_k\sum_{l=0}^{k}\,s(n,k-l)\,s(m,l).
\end{eqnarray}
The first few factorial cumulants are
\begin{eqnarray}\label{eq:factorial-cumulants-3}
    F^{(1,0)}&=&\alpha_B\,\nbcan,\nonumber\\
    F^{(0,1)}&=&\alpha_{\bar B}\,\nabcan,\nonumber\\
    F^{(2,0)}&=&(\alpha_B)^2\,\big(c_2-\nbcan\big),\nonumber\\
    F^{(1,1)}&=&\alpha_B\,\alpha_{\bar B}\,c_2,\nonumber\\
    F^{(0,2)}&=&(\alpha_{\bar B})^2\,\big(c_2-\nabcan\big),\\
    F^{(3,0)}&=&(\alpha_B)^3\,\big(c_3-3\,c_2+2\,\nbcan\big),\nonumber\\
    F^{(2,1)}&=&(\alpha_B)^2\,\alpha_{\bar B}\,\big(c_3-c_2\big),\nonumber\\
    F^{(1,2)}&=&\alpha_B\,(\alpha_{\bar B})^2\,\big(c_3-c_2\big),\nonumber\\
    F^{(0,3)}&=&(\alpha_{\bar B})^3\,\big(c_3-3\,c_2+2\,\nabcan\big).\nonumber
\end{eqnarray}\\

\subsection{Cumulants of \texorpdfstring{$\delta N_B$}{d N B} and \texorpdfstring{$\delta N_{\bar B}$}{d N B}}

The generating function for cumulants of $\delta N_B$ and $\delta N_{\bar B}$ in the subsystem is obtained from (\ref{eq:generating-function-fact-cum}) by the replacements $x\to e^t$ and $\bar{x}\to e^s$,
\begin{eqnarray}\label{eq:generating-function-NB-cum}
    g_{c}(t,s)&=&\frac{B}{2}\big[\ln\big(q(t)\big)-\ln\big(\bar{q}(s)\big)\big]\\
    &+&\ln\big[I_B\big(2\,z\,\sqrt{q(t)\,\bar{q}(s)}\big)\big],\nonumber
\end{eqnarray}
where $q(t)=1-\alpha_B+\alpha_B\,e^t$ and $\bar{q}(s)=1-\alpha_{\bar B}+\alpha_{\bar B}\,e^s$. 

The cumulants in the subsystem $A$, defined by the acceptance probabilities $\alpha_B$ and $\alpha_{\bar B}$,
are given by
\begin{equation}
    C^{(n,m)}=\frac{\partial^n}{\partial t^n}\,\frac{\partial^m}{\partial s^m}\,g_c(t,s)|_{t=s=0}.
\end{equation}

As shown in \ref{sect:App-c}, it follows from the relation between the generating functions, $g_c(t,s)=g_{fc}(e^t,e^s)$, that the cumulants can be obtained from the corresponding factorial cumulants using
\begin{equation}\label{eq:cumulant-fac-cumulant-relation-subsystem-text}
    C^{(n,m)}=\sum_{k_1=0}^n\,\sum_{k_2=0}^m\,F^{(k1,k2)}\,S(n,k_1)\,S(m,k_2).
\end{equation}
Now, inserting the expression for the factorial cumulants in the canonical ensemble (\ref{eq:factorial-cumulants-closed-form}) in (\ref{eq:cumulant-fac-cumulant-relation-subsystem-text}), we find
\begin{eqnarray}\label{eq:baryon-antib-cumulants-sub}
    C^{(n,m)}&=&\frac{B}{2}\,\Big(\kappa_{\rm Ber}^{(n)}(\alpha_B)\,\delta_{m,0}-\kappa_{\rm Ber}^{(m)}(\alpha_{\bar B})\,\delta_{n,0}\Big)\\
    &+&\sum_{k=1}^{n+m}c_k\sum_{i=0}^k B_{n,k-i}\big(\kappa_{\rm Ber}^{(1)}(\alpha_B),\kappa_{\rm Ber}^{(2)}(\alpha_B),\dots\big)\nonumber\\
    &\times&B_{m,i}\big(\kappa_{\rm Ber}^{(1)}(\alpha_{\bar B}),\kappa_{\rm Ber}^{(2)}(\alpha_{\bar B}),\dots\big),\nonumber
\end{eqnarray}
where \cite{Braun-Munzinger:2020jbk}
\begin{equation}\label{eq:Bernoulli-cumulant}
    \kappa_{\rm Ber}^{(n)}(\alpha)=\delta_{n,1}+(-1)^{1+n}\,{\rm Li}_{1-n}(1-1/\alpha),
\end{equation}
are the cumulants of the Bernoulli distribution with success probability $p$ and ${\rm Li}_{n}(x)$ is the polylogarithm. 
In obtaining (\ref{eq:baryon-antib-cumulants-sub}) we used the relations
\begin{eqnarray}
&&\sum_{k=1}^n S(n,k)\,(-\alpha)^k\,(k-1)!=-\kappa^{(n)}_{\rm Ber}(\alpha),\\
&&\sum_{k=1}^n S(n,k)\,(\alpha)^k\,s(k,i)=B_{n,i}\big(\kappa_{\rm Ber}^{(1)}(\alpha),\kappa_{\rm Ber}^{(2)}(\alpha),\dots\big).\nonumber
\end{eqnarray}
We provide explicit expressions for the first few cumulants in the subsystem,
\begin{eqnarray}\label{eq:cumulants-3}
    C^{(1,0)}&=&\alpha_B\,\nbcan,\nonumber\\
    C^{(0,1)}&=&\alpha_{\bar B}\,\nabcan,\nonumber\\
    C^{(2,0)}&=&(\alpha_B)^2\,c_2+\alpha_B(1-\alpha_B)\,\nbcan,\nonumber\\
    C^{(1,1)}&=&\alpha_B\,\alpha_{\bar B}\,c_2,\nonumber\\
    C^{(0,2)}&=&(\alpha_{\bar B})^2\,c_2+\alpha_{\bar B}(1-\alpha_{\bar B})\,\nabcan,\\
    C^{(3,0)}&=&(\alpha_B)^3\,c_3+3(\alpha_B)^2\,(1-\alpha_B)\,c_2\nonumber\\
             &+&\alpha_B(1-3\,\alpha_B+2\,(\alpha_B)^2)\,\nbcan,\nonumber\\
    C^{(2,1)}&=&(\alpha_B)^2\,\alpha_{\bar B}\,c_3+\alpha_B\,\alpha_{\bar B}\,(1-\alpha_B)\,c_2,\nonumber\\
    C^{(1,2)}&=&\alpha_B\,(\alpha_{\bar B})^2\,c_3+\alpha_B\,\alpha_{\bar B}\,(1-\alpha_{\bar B})\,c_2,\nonumber\\
    C^{(0,3)}&=&(\alpha_{\bar B})^3\,c_3+3(\alpha_{\bar B})^2\,(1-\alpha_{\bar B})\,c_2\nonumber\\
             &+&\alpha_{\bar B}(1-3\,\alpha_{\bar B}+2\,(\alpha_{\bar B})^2)\,\nabcan.\nonumber
\end{eqnarray}

\subsection{Cumulants of the net baryon number}

The generating function for net baryon number cumulants  in a subsystem, is obtained by making the substitutions $x\to e^t$ and $\bar{x}\to e^{-t}$ in (\ref{eq:generating-function-fact-cum}). The generating function can also be derived starting from the probability distribution $P_A(B_A)$ of the net baryon number in the subsystem~\cite{Bzdak:2012an,Braun-Munzinger:2020jbk}. One finds\footnote{The generating function for cumulants of $\delta(N_B-N_{\bar B})$ in a finite acceptance, $g_{net}$, reduces to the generating function for cumulants of $\delta(N_B+N_{\bar B})$ in the full system, Eq.~(\ref{eq:can_cumulant_gen_function}), in the limit $\alpha_B\to 1$ and $\alpha_{\bar B}\to -e^t$.}
\begin{eqnarray}\label{eq:generating-function}
    g_{net}(t)&=&\ln\left(\sum_{B_A} P_A(B_A)e^{B_A t}\right)\nonumber\\
    &=&\frac{B}{2}\left[\ln(q_1(t))-\ln(q_2(t))\right]+\ln\left\{I_B[2\, z\, \sqrt{q_1(t)\, q_2(t)}]\right\},
\end{eqnarray}
where $B$ is the total net baryon number, $q_1(t)=1-\alpha_B+\alpha_B\, e^t$ and $q_2(t)=1-\alpha_{\bar B}+\alpha_{\bar B}\,e^{-t}$.

In order to compute the net baryon number cumulants, we employ the formula of Fa\`{a} di Bruno (\ref{eq:faa-di-bruno-1})
to compute derivatives of $g_{net}(t)$.
For the first two terms in (\ref{eq:generating-function}), which we denote by $(a)$ and $(b)$, we choose $f(y)=\ln(y)$, $h(t)=q_1(t)$ and $f(y)=-\ln(y)$, $h(t)=q_2(t)$, as in~\cite{Braun-Munzinger:2020jbk}. Employing again the Fa\`{a} di Bruno formula for functions of the form $\log(f(y))$~\cite{Comtet:1974}, we find a closed-form expression for the corresponding contribution to the net baryon number cumulants~\cite{Braun-Munzinger:2020jbk}
\begin{eqnarray}\label{eq:cumulants-a-b}
    \kappa^{(a+b)}_n&=&\frac{B}{2}\Big(\kappa_{\rm Ber}^{(n)}(\alpha_B)+(-1)^{(n+1)}\kappa_{\rm Ber}^{(n)}(\alpha_{\bar B})\Big)\nonumber\\
    &\equiv&B\,k^{(n)}_+,
\end{eqnarray}
where $\kappa_{\rm Ber}^{(n)}(\alpha_B)$ is the $n$:th cumulant of the Bernoulli distribution (\ref{eq:Bernoulli-cumulant}) and the second line defines $k^{(n)}_+$.

Now, in the evaluation of the derivatives of the last term in (\ref{eq:generating-function}), denoted by ${(c)}$, we deviate slightly from ~\cite{Braun-Munzinger:2020jbk}, and choose $f(y)=\ln\big(I_B[2\,z\,e^y]\big)$ and $h(t)=\frac12\big(\ln(q_1(t))+\ln(q_2(t))\big)$ and evaluate the derivatives at $y=1$ and $t=0$, respectively. Using the fact that the derivatives of $f(y)$ are given by the cumulants $C_k$ (cf. (\ref{eq:can_cumulant_gen_function},\ref{eq:can-cumulant-Ntot})) and those of $h(t)$ by~\cite{Braun-Munzinger:2020jbk}
\begin{equation}
    h^{(n)}\equiv k_-^{(n)}=\frac12\Big(\kappa_{\rm Ber}^{(n)}(\alpha_B)-(-1)^{(n+1)}\kappa_{\rm Ber}^{(n)}(\alpha_{\bar B})\Big),
\end{equation}
we find that the corresponding contributions to the cumulants are given by
\begin{equation}\label{eq:kappa-c-cumulants}
    \kappa^{(c)}_n=\sum_{k=1}^n\,C_k\,B_{n,k}(k_-^{(1)},\dots ,k_-^{(n-k+1)}).
\end{equation}
Thus, the net baryon cumulants in a subsystem are closely related to the cumulants of $\delta N^{tot}=\delta (N_B+N_{\bar B})$ in the full canonical system, where the net baryon number is strictly conserved. Collecting all terms, we find a compact form for the net baryon cumulants in a subsystem of a canonical system,
\begin{equation}\label{eq:cumulants-closed-form}
    \kappa_n=B\,k_+^{(n)}+\sum_{k=1}^n\,C_k\,B_{n,k}(k_-^{(1)},\dots ,k_-^{(n-k+1)}).
\end{equation}
We note that, using (\ref{eq:Nn-Ck-relation}), the cumulants $\kappa_n$ can be expressed in terms of the factorial cumulants $F_k$ rather than cumulants $C_k$. The resulting analytic expression for $\kappa_n$ is identical to the one obtained in~\cite{Braun-Munzinger:2020jbk} in terms of the functions $f^{(k)}(z)$, which as noted above, are equal to the factorial cumulants (\ref{eq:factorial-cumulants-tot}).

In \ref{sect:App-c}, we obtain general relations of the net baryon cumulants to the baryon and antibaryon cumulants in the subsystem,
\begin{equation}\label{eq:kappa-Cmn-text}
    \kappa_n=\sum_{i=0}^n\,\binom{n}{i}\,(-1)^{n-i}\,C^{(i,n-i)},
\end{equation}
where $\binom{n}{i}$ is a binomial coefficient, and to the factorial ones,
\begin{equation}\label{eq:cumulant-factorial-cumulant-relation}
    \kappa_n=\sum_{\substack{k_1,k_2\\ k_1+k_2\geq 0}}^{n}F^{(k_1,k_2)}\,\sum_{i=0}^n\binom{n}{i}\,(-1)^{n-i}\,S(i,k_1)\,S(n-i,k_2).
\end{equation}
One thus finds,
\begin{eqnarray}\label{eq:cumulants-from-can-cumulants}
    \kappa_1&=&F^{(1,0)}-F^{(0,1)},\nonumber\\
    \kappa_2&=&F^{(1,0)}+F^{(0,1)}+F^{(2,0)}+F^{(0,2)}-2\,F^{(1,1)},\nonumber\\
    \kappa_3&=&F^{(1,0)}-F^{(0,1)}+3(F^{(2,0)}-F^{(0,2)})\nonumber\\
    &-&3(F^{(2,1)}-F^{(1,2)})+F^{(3,0)}-F^{(0,3)},\nonumber\\
    \kappa_4&=&F^{(1,0)}+F^{(0,1)}+7(F^{(2,0)}+F^{(0,2)})+6(F^{(3,0)}+F^{(0,3)})\nonumber\\
    &-&6(F^{(2,1)}+F^{(1,2)})-4(F^{(3,1)}+F^{(1,3)})+6\,F^{(2,2)}-2\,F^{(1,1)}\nonumber\\
    &+&F^{(4,0)}+F^{(0,4)},\\
    \kappa_5&=&F^{(1,0)}-F^{(0,1)}+15(F^{(2,0)}-F^{(0,2)})+25(F^{(3,0)}-F^{(0,3)})\nonumber\\
    &+&10(F^{(4,0)}-F^{(0,4)})-15(F^{(2,1)}-F^{(1,2)})-20(F^{(3,1)}-F^{(1,3)})\nonumber\\
    &-&5(F^{(4,1)}-F^{(1,4)})+10(F^{(3,2)}-F^{(2,3)})+F^{(5,0)}-F^{(0,5)}\nonumber
\end{eqnarray}
\begin{eqnarray}
    \kappa_6&=&F^{(1,0)}+F^{(0,1)}+31(F^{(2,0)}+F^{(0,2)})+90(F^{(3,0)}+F^{(0,3)})\nonumber\\
    &+&65(F^{(4,0)}+F^{(0,4)})+15(F^{(5,0)}+F^{(0,5)})-30(F^{(2,1)}+F^{(1,2)})\nonumber\\
    &-&80(F^{(3,1)}+F^{(1,3)})-45(F^{(4,1)}+F^{(1,4)})-6(F^{(5,1)}+F^{(1,5)})\nonumber\\
    &+&30(F^{(3,2)}+F^{(2,3)})+15(F^{(4,2)}+F^{(2,4)})-20F^{(3,3)}+30F^{(2,2)}\nonumber\\
    &-&2F^{(1,1)}+F^{(6,0)}+F^{(0,6)},\nonumber
\end{eqnarray}
When the factorial cumulants (\ref{eq:factorial-cumulants-closed-form}, \ref{eq:factorial-cumulants-3}) are plugged into (\ref{eq:cumulant-factorial-cumulant-relation},\ref{eq:cumulants-from-can-cumulants}), one recovers 
the explicit expressions given in~\cite{Braun-Munzinger:2020jbk}
\section{High- and low-energy limits}

In the high-energy limit, ${\rm max}(B,1/2)<<2\,z$, one finds~\cite{Braun-Munzinger:2020jbk}
\begin{equation}\label{eq:series-exp-fn}
    F_n=2\,z\,\delta_{n,1}-(-1)^{n-1}\,(n-1)!\Big(\frac12 -n\frac{4\,B^2-1}{16\,z}\Big)+\mathcal{O}(1/z^2).
\end{equation}
Inserting this into (\ref{eq:Nn-Ck-relation}), one finds the high-energy limit of the cumulants of $\delta(N_B+N_{\bar B})$,
\begin{equation}\label{eq:N_to_f_rel_he}
    C_m^{(HE)}=\sum_{n=1}^m\,S(m,n)\,\left(2\,z\,\delta_{n,1}-(-1)^{n-1}\,(n-1)!\Big(\frac12 -n\frac{4\,B^2-1}{16\,z}\Big)\right),
\end{equation}
where terms of order $\mathcal{O}(1/z^2)$ and higher have been dropped. Using the relations
\begin{eqnarray}
    \sum_{n=1}^m\,S(m,n)\,(-1)^{n-1}\,(n-1)!&=&\delta_{m,1},\\
    \sum_{n=1}^m\,S(m,n)\,(-1)^{n-1}\,n!&=&(-1)^{m-1},
\end{eqnarray}
we find
\begin{eqnarray}\label{eq:Nk-high-energy}
    C_1^{(HE)}&=&2\,z-\frac12+\frac{4\,B^2-1}{16\,z},\\
    C_n^{(HE)}&=&2\,z+(-1)^{n-1}\,\frac{4\,B^2-1}{16\,z}\qquad ({\rm for}\,\,n>1).\nonumber
\end{eqnarray}
When the cumulants (\ref{eq:Nk-high-energy}) are plugged into (\ref{eq:cumulants-closed-form}), one recovers the high-energy limits of $\kappa_n$ derived in~\cite{Braun-Munzinger:2020jbk}.

We note that for $n>1$ and $z\to\infty$, the cumulants 
\begin{equation}
C_n^{(HE)}=2\,z=C_1^{(HE)}+\frac12    
\end{equation}
are independent of $n$, like the cumulants of a Poisson distribution. Thus, in the canonical ensemble, the high-energy limit of the cumulants of $N^{(tot)}$ are, except for the first one, equal to those of a Poisson distribution with the cumulants $\langle N_B+N_{\bar B}\rangle+1/2$. Consequently, in the high-energy limit ($B/z\to 0$), the fluctuations of $\delta(N_B+N_{\bar B})$ in the canonical ensemble are, apart from a small correction, the same as those in the grand-canonical ensemble.

In the low-energy limit, we expand the particle numbers about $z=0$.
For $B>0$, one finds\footnote{Compared to the result given in~\cite{Braun-Munzinger:2020jbk}, we have here included the $z^4$ terms. }
\begin{eqnarray}
    \langle N_B\rangle &=&B+\frac{z^2}{B+1}-\frac{z^4}{(B+1)^2(B+2)}+\mathcal{O}(z^6/B^5),\\
    \langle N_{\bar B}\rangle
    &=& \frac{z^2}{B+1}-\frac{z^4}{(B+1)^2(B+2)}+\mathcal{O}(z^6/B^5).\label{eq:low-energy-multiplicities}
\end{eqnarray}
Thus, retaining terms of order $z^4$,
\begin{equation}
    C_1=F_1=\langle N_B+N_{\bar B}\rangle\approx B+\frac{2\,z^2}{B+1}-\frac{2\,z^4}{(B+1)^2(B+2)}.
\end{equation}
Using the recurrence relation 
for $F_{n}$
(\ref{eq:rec-relation-F}), one finds for the low-energy factorial cumulants
\begin{eqnarray}
    F_n&=&(-1)^{n-1}\,(n-1)!\,B+(\delta_{n,1}+\delta_{n,2})\,\frac{2\,z^2}{B+1}\\
    &-&(\delta_{n,1}+3\,\delta_{n,2}+6\,\delta_{n,3}+6\,\delta_{n,4})\,\frac{2\,z^4}{(B+1)^2(B+2)}.\nonumber
\end{eqnarray}
Similarly, using the recurrence relation for $C_k$ (\ref{eq:rec-rel-N}), one finds~\footnote{Note that both the high- and low-energy limits of the cumulants and factorial cumulants satisfy the relations (\ref{eq:Cn_to_Nk-relation}) and (\ref{eq:Nn-Ck-relation}).}
\begin{equation}\label{eq:Nk-low-energy}
    C_k^{(LE)}=\frac{2^{k}\, z^2}{B+1} -\frac{ 2^{2\,k-1}\,z^4}{(B+1)^2(B+2)},\quad (k>1).
\end{equation}

For the cumulants of the baryon number one then finds
\begin{eqnarray}
    c_1^{(LE)}&=&B/2+\frac{z^2}{B+1}-\frac{z^4}{(B+1)^2(B+2)}=\frac12\langle N_{B}+ N_{\bar B}\rangle,\\
    c_k^{(LE)}&=&\frac{z^2}{B+1} -\frac{2^{k-1}\,z^4}{(B+1)^2(B+2)}= \langle N_{\bar B}\rangle +\mathcal{O}(z^4), \quad (k>1).\label{eq:LE-baryon-cumulant}
\end{eqnarray}
Thus, in the low-energy limit $(z\to 0)$, the fluctuations of the baryon number, $c_k=\langle \left(\delta N_B\right)^k\rangle_c\,\,\,(k>1)$, approach the cumulants of a Poisson distribution with expectation value  $\langle N_{\bar B}\rangle$. 
In other words, in the canonical ensemble, fluctuations of the baryon number 
are at low energies equal to the fluctuations of the antibaryon number in the grand canonical ensemble.

\begin{figure}[t]
    \centering
    \includegraphics[width=.48\linewidth,clip=true]{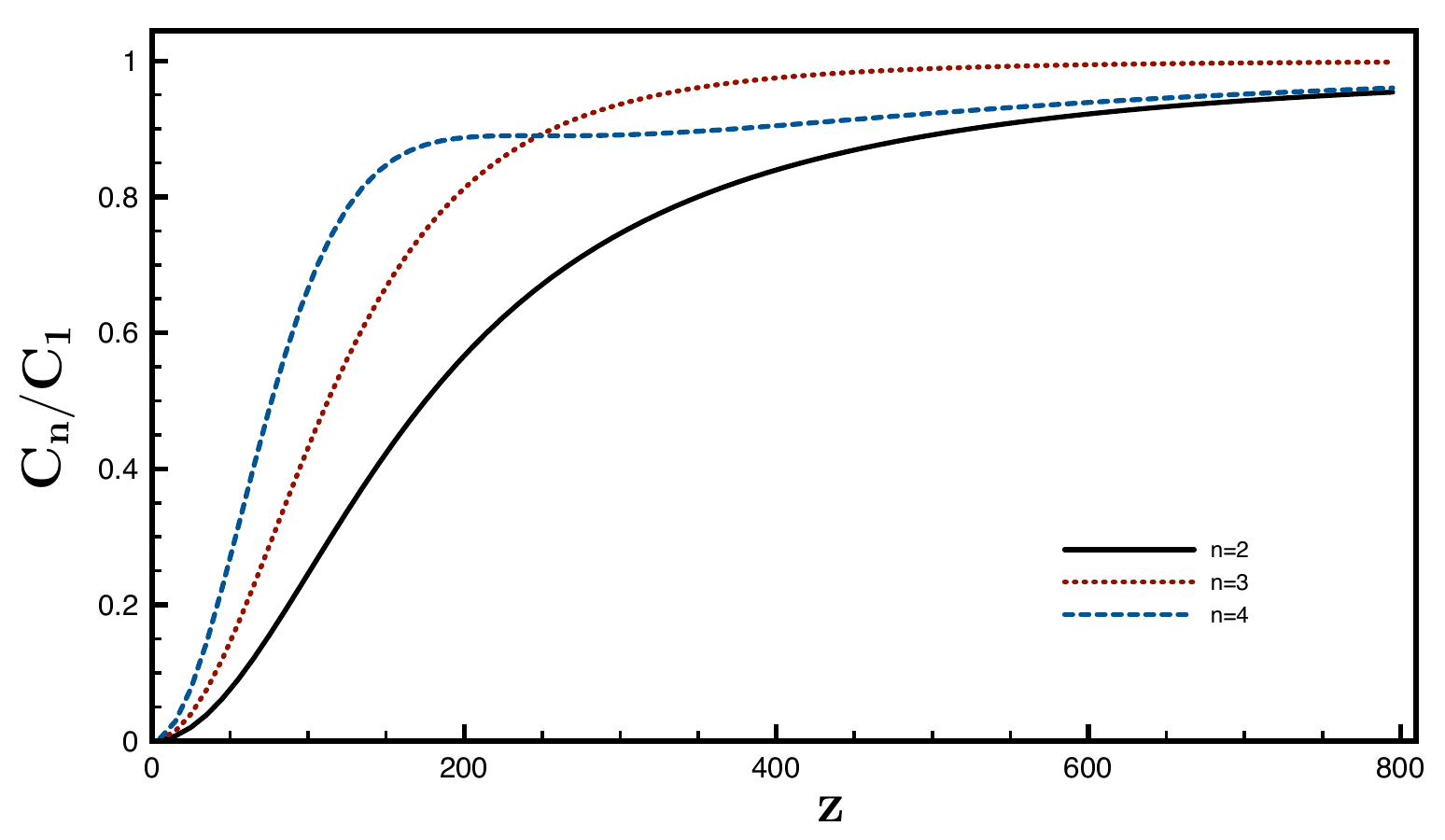}
    \includegraphics[width=.48\linewidth,clip=true]{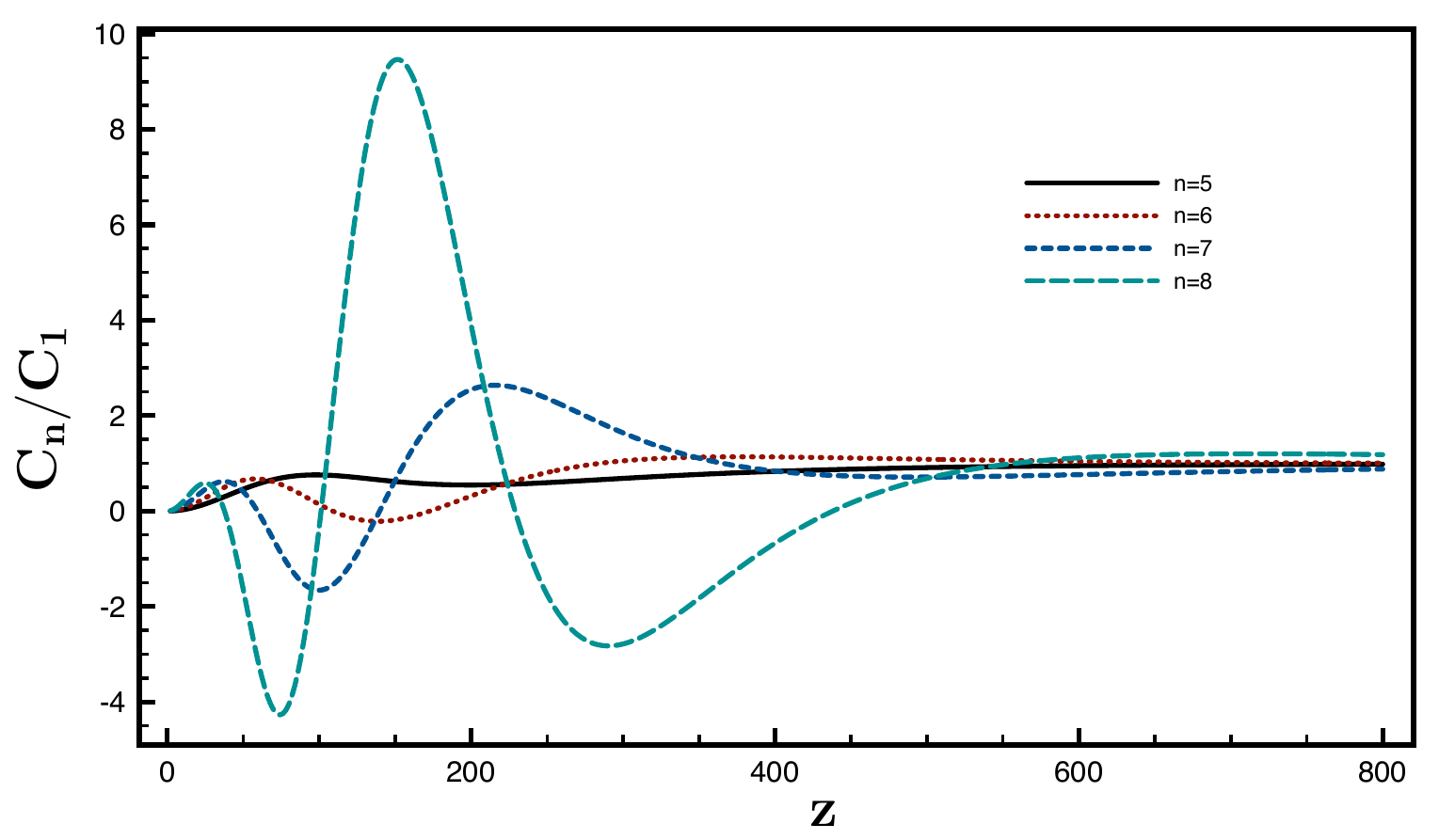}
    \caption{The cumulants of $\delta(N_B+N_{\bar B})$, $C_n$, normalized to $N_B+N_{\bar B}=C_1$. The corresponding beam energy increases with increasing $z$.}
    \label{fig:baryon-antibaryon-cumulants}
\end{figure}

\begin{figure}[t]
    \centering
    \includegraphics[width=.48\linewidth,clip=true]{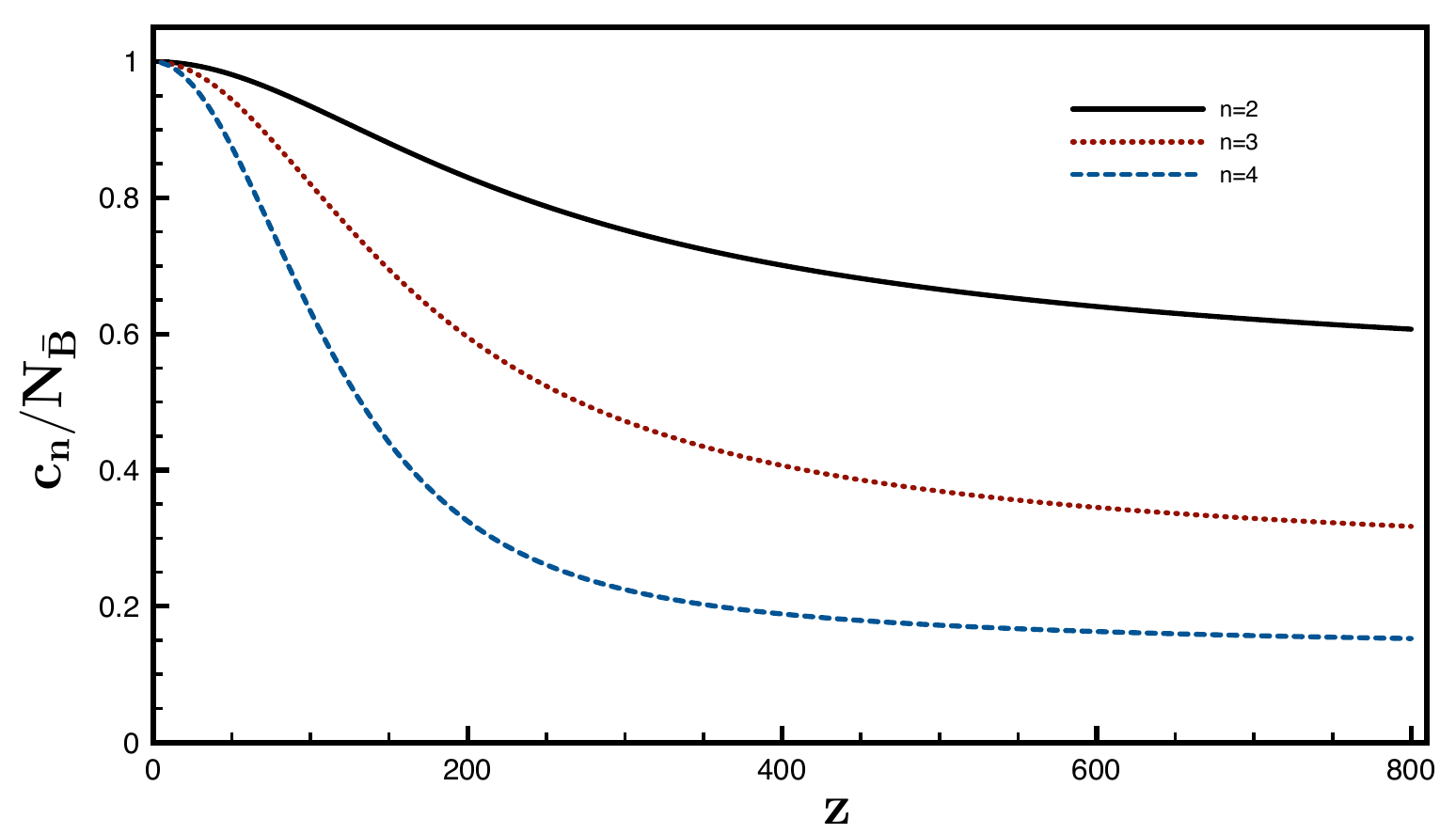}
    \includegraphics[width=.48\linewidth,clip=true]{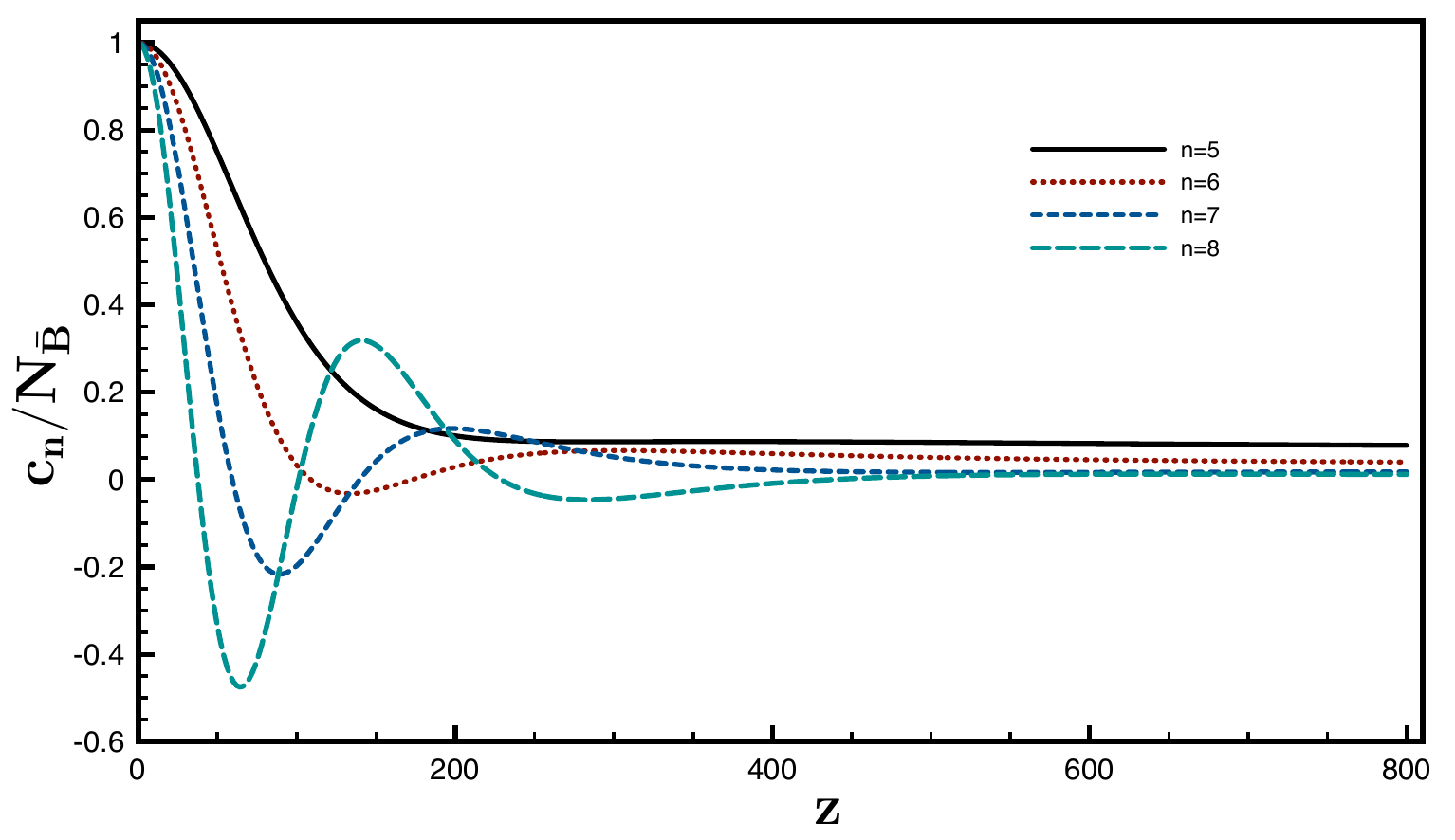}
    \caption{The cumulants of $\delta(N_B)$, $c_n$, normalized to $N_{\bar B}$.}
    \label{fig:baryon-cumulants}
\end{figure}

The cumulants of $\delta(N_B+N_{\bar B})$, (\ref{eq:cumulants-Ntot}), are shown in Fig.~\ref{fig:baryon-antibaryon-cumulants} as functions of $z$ for net baryon number $B=350$.
In Fig.~\ref{fig:baryon-cumulants} the corresponding cumulants of $\delta N_B$, (\ref{eq:shorthand-2}), are shown. The behaviour in the high- and low-energy limits (large and small $z$, respectively) of the cumulants shown in Figs.~\ref{fig:baryon-antibaryon-cumulants} and \ref{fig:baryon-cumulants} is consistent with the analytic results (\ref{eq:Nk-high-energy}) and (\ref{eq:Nk-low-energy}).

\section{Summary and conclusions} 
The main objective of these studies was to calculate fluctuations of the multiplicity distributions of baryons, antibaryons, as well as their sum and difference in a canonical ensemble. 
We have focused on cumulants and factorial cumulants, assuming that   baryons are  correlated  by exact conservation of the net baryon number in the  full system. 
To this end we have adopted a statistical model of baryons and antibaryons in the spirit of the S-matrix approach,  where  to leading order in the fugacity expansion, the  partition function takes the form of an ideal gas.    

We have derived analytic expressions 
for the cumulants and factorial cumulants of the baryon and antibaryon numbers in the canonical ensemble. Moreover, based on the generalized Fa\`{a} di Bruno formula for computing the derivatives of composite functions, we obtained closed-form results for cumulants and factorial cumulants of arbitrary order in a subsystem constrained
by exact conservation of the net baryon number in the full system. 
We also presented general relations, independent of assumptions on the partition function, between cumulants and factorial cumulants as well as between net-baryon-number fluctuations in a subsystem and those of the baryon and antibaryon numbers of the full system. Finally, we deduced the asymptotic forms of the cumulants and factorial cumulants of baryon and antibaryon multiplicities in the high- and low-energy limits.

The analytic results for fluctuation observables of the baryon and antibaryon numbers in a subsystem,  given a canonical partition function for the full system, as well as the general relations between observables, provide a firm baseline for the phenomenological interpretation of  fluctuation data obtained in heavy-ion collisions.  Applications of these results will be presented in an  upcoming publication.

\section{Acknowledgements}
We acknowledge stimulating discussions with Peter Braun-Munzinger, Anar Rustamov and Johanna Stachel.   K.R. also acknowledges  the supported by the Polish National Science Center (NCN) under OPUS Grant No. 2018/31/B/ST2/01663,  and by the Polish Ministry of Science.

\appendix
\section{Multivariate Bell polynomials}\label{sec:app-a}

In order to express the derivatives of a composite function of several variables, like $g_{fc}(x,\bar{x})$ in (\ref{eq:generating-function-fact-cum}), in a form analogous to the Fa\`{a} di Bruno formula one needs a generalization of the Bell polynomials~\cite{Bell:1927}.

The multivariate partial Bell polynomials can be obtained from a generalization of the generating function for the standard partial Bell polynomials~\cite{Schumann:2019xy}
\begin{equation}\label{eq:gen-function-Bell}
    \Phi(t,u)=\exp\left(u\sum_j x^{(j)}\frac{t^j}{j!}\right)=\sum_{n\geq k\geq 0}B_{n,k}(x^{(1)},x^{(2)},\dots,x^{(n-k+1)})\,\frac{t^n}{n!}\,u^k.
\end{equation}
The regular partial Bell polynomials are given by
\begin{equation}\label{eq:partial-Bell}
    B_{n,k}(x^{(1)},x^{(2)},\dots,x^{(n-k+1)})=\frac{1}{k!}\frac{\partial^n}{\partial t^n}\,\frac{\partial^k}{\partial u^k}\,\Phi(t,u)|_{t=u=0}.
\end{equation}
The Bell polynomials enter the Fa\`{a} di Bruno formula (\ref{eq:faa-di-bruno-1}) for the derivatives of a composite function $f(g(z))$,
Below we present the generalizations of (\ref{eq:faa-di-bruno-1}) to composite functions of the form $f(g(z_1,z_2))$ and $f(g_1(z),g_2(z))$.

In Ref.~\cite{Schumann:2019xy} the generating function for the general case of arbitrary dimensions of $z$ and $g$ is given. In order to keep the notation transparent, we give the generating function for two-dimensional $\vec{z}=(z_1,z_2)$ and $\vec{g}=(g_1,g_2)$,
\begin{eqnarray}\label{eq:gen-function-multiv-Bell}
    \Phi(t_1,t_2,u_1,u_2)&=&\exp\left(\sum_{\substack{j_1,j_2\\ j_1+j_2>0}}^\infty \big[u_1*x_1^{(j_1,j_2)}+u_2*x_2^{(j_1,j_2)}\big]\frac{t_1^{j_1}\,t_2^{j_2}}{j_1!\,j_2!}\right)\\
    &=&\sum_{\substack{n_1,n_2,k_1,k_2\\ n_1+n_2\geq k_1+k_2\geq 0}}B_{\{n_1,n_2\},\{k_1,k_2\}}\big(\{\vec{x}^{(i,j)}\}\big)\,\frac{t_1^{n_1}\,t_2^{n_2}}{n_1!\,n_2!}\,(u_1^{k_1}+u_2^{k_2}).\nonumber
\end{eqnarray}
The generalized Bell polynomials are then given by
\begin{eqnarray}\label{eq:gen-Bell-22}
    B_{n_1,n_2;k_1,k_2}(\{\vec{x}^{(i,j)}\})&=&\frac{1}{k_1!\,k_2!}\\
    &\times&\frac{\partial^{n_1}}{\partial t_1^{n_1}}\frac{\partial^{n_2}}{\partial t_2^{n_2}}\frac{\partial^{k_1}}{\partial u_1^{k_1}}\frac{\partial^{k_2}}{\partial u_2^{k_2}}\,\Phi(t_1,t_2,u_1,u_2)|_{t_1=t_2=u_1=u_2=0}.\nonumber
\end{eqnarray}

Now, for computing the factorial cumulants, we need the case with vector argument $\vec{z}=(z_1,z_2)$ and a scalar function $g(z_1,z_2)$,
\begin{eqnarray}\label{eq:gen-function-multiv-Bell-21}
    \Phi(t_1,t_2,u)&=&\exp\left(u\,\sum_{\substack{j_1,j_2\\ j_1+j_2>0}}^\infty x^{(j_1,j_2)}\frac{t_1^{j_1}\,t_2^{j_2}}{j_1!\,j_2!}\right)\\
    &=&\sum_{\substack{n_1,n_2,k\\ n_1+n_2\geq k\geq 0}}B_{n_1,n_2;k}\big(\{x^{(i,j)}\}\big)\,\frac{t_1^{n_1}\,t_2^{n_2}}{n_1!\,n_2!}\,u^{k}.\nonumber
\end{eqnarray}
The corresponding partial Bell polynomials are given by,
\begin{eqnarray}\label{eq:gen-Bell-21}
    B_{n_1,n_2;k}(\{x^{(i,j)}\})&=&\frac{1}{k!}\,\frac{\partial^{n_1}}{\partial t_1^{n_1}}\frac{\partial^{n_2}}{\partial t_2^{n_2}}\frac{\partial^{k}}{\partial u^{k}}\,\Phi(t_1,t_2,u)|_{t_1=t_2=u=0}.
\end{eqnarray}
The first few generalized Bell polynomials are
\begin{eqnarray}\label{eq:app-7}
    B_{1,0;1}(\{x^{(i,j)}\})&=&x^{(1,0)}\,,\nonumber\\
    B_{0,1;1}(\{x^{(i,j)}\})&=&x^{(0,1)}\,,\nonumber\\
    B_{2,0;1}(\{x^{(i,j)}\})&=&x^{(2,0)}\,,\nonumber\\
    B_{2,0;2}(\{x^{(i,j)}\})&=&(x^{(1,0)})^2\,,\\
    B_{0,2;1}(\{x^{(i,j)}\})&=&x^{(0,2)}\,,\nonumber\\
    B_{0,2;2}(\{x^{(i,j)}\})&=&(x^{(0,1)})^2\,,\nonumber\\
    B_{1,1;1}(\{x^{(i,j)}\})&=&x^{(1,1)}\,,\nonumber\\
    B_{1,1;2}(\{x^{(i,j)}\})&=&x^{(1,0)}\,x^{(0,1)}\,,\nonumber\\
    B_{3,0;1}(\{x^{(i,j)}\})&=&x^{(3,0)}\,,\nonumber\\
    B_{3,0;2}(\{x^{(i,j)}\})&=&3\,x^{(1,0)}\,x^{(2,0)}\,,\nonumber\\
    B_{3,0;3}(\{x^{(i,j)}\})&=&(x^{(1,0)})^3\,,\nonumber\\
    B_{2,1;1}(\{x^{(i,j)}\})&=&x^{(2,1)}\,,\nonumber\\
    B_{2,1;2}(\{x^{(i,j)}\})&=&2\,x^{(1,0)}\,x^{(1,1)}+x^{(0,1)}\,x^{(2,0)},\nonumber\\
    B_{2,1;3}(\{x^{(i,j)}\})&=&x^{(0,1)}\,(x^{(1,0)})^2.\nonumber\\
    B_{1,2;1}(\{x^{(i,j)}\})&=&x^{(1,2)}\,,\nonumber\\
    B_{1,2;2}(\{x^{(i,j)}\})&=&2\,x^{(0,1)}\,x^{(1,1)}+x^{(1,0)}\,x^{(0,2)},\nonumber\\
    B_{1,2;3}(\{x^{(i,j)}\})&=&x^{(1,0)}\,(x^{(0,1)})^2.\nonumber\\
    B_{0,3;1}(\{x^{(i,j)}\})&=&x^{(0,3)}\,,\nonumber\\
    B_{0,3;2}(\{x^{(i,j)}\})&=&3\,x^{(0,1)}\,x^{(0,2)}\,,\nonumber\\
    B_{0,3;3}(\{x^{(i,j)}\})&=&(x^{(0,1)})^3\,,\nonumber
\end{eqnarray}

A combinatorial interpretation of the Bell polynomial $B_{n,m;k}({x^{\{i,j\}}})$ goes as follows. Consider a collection of $n$ blue beads and $m$ red ones. How can these be split into $k$ groups is encoded in the Bell polynomials. A group consisting of $i$ blue beads and $j$ red ones is denoted by $x^{\{i,j\}}$. For instance, a system consisting of two blue beads and one red one can be split into two groups in three ways. Two with one blue and one red bead in one group and the remaining blue bead in the other group and one with the two blue beads in one group and the red one in the other. This case corresponds to the Bell polynomial $B_{2,1;2}(\{x^{(i,j)}\})$ in (\ref{eq:app-7}).

The generalized Fa\`{a} di Bruno formula for computing the derivatives of a composite function of two variables, $f(g(z_1,z_2))$, needed for the calculation of the factorials cumulants is then,
\begin{eqnarray}\label{eq:faa-di-bruno-mult1}
    \frac{\partial^{n_1}}{\partial z_1^{n_1}}\,\frac{\partial^{n_2}}{\partial z_2^{n_2}}f(g(z_1,z_2))=\sum_{k=1}^{n_1+n_2}f^{(k)}\,B_{n_1,n_2;k}(\{g^{(i,j)}\}).
\end{eqnarray}
The first few terms are
\begin{eqnarray}
    \frac{\partial}{\partial z_1}\,f(g(z_1,z_2))&=&f^{(1)}g^{(1,0)}\,,\nonumber\\
    \frac{\partial}{\partial z_2}\,f(g(z_1,z_2))&=&f^{(1)}g^{(0,1)}\,,\nonumber\\
    \frac{\partial^2}{\partial z_1^2}\,f(g(z_1,z_2))&=&f^{(1)}g^{(2,0)}+f^{(2)}\big(g^{(1,0)}\big)^2\,,\\
    \frac{\partial^2}{\partial z_2^2}\,f(g(z_1,z_2))&=&f^{(1)}g^{(0,2)}+f^{(2)}\big(g^{(0,1)}\big)^2\,,\nonumber\\
    \frac{\partial}{\partial z_1}\frac{\partial}{\partial z_2}\,f(g(z_1,z_2))&=&f^{(1)}g^{(1,1)}+f^{(2)}\,g^{(1,0)}\,g^{(0,1)}\,.\nonumber
\end{eqnarray}

We also need the Bell polynomials for computing the derivatives of a function of the form $f(g_1(z),g_2(z))$. The generating function is:
\begin{eqnarray}
    \Phi(t,u_1,u_2)&=&\exp\left(\sum_{j=1}^\infty \big[u_1*x_1^{(j)}+u_2*x_2^{(j)}\big]\frac{t^{j}}{j!}\right)\\
    &=&\sum_{\substack{n,k_1,k_2\\ n\geq k_1+k_2\geq 0}}B_{n;k_1,k_2}\big(\{\vec{x}^{(i)}\}\big)\,\frac{t^{n}}{n!}\,(u_1^{k_1}+u_2^{k_2}).\nonumber
\end{eqnarray}
and the corresponding generalized Bell polynomials are given by,
\begin{equation}
    B_{n;k_1,k_2}(\{\vec{x}^{(i)}\})=\frac{1}{k_1!\,k_2!}
    \frac{\partial^{n}}{\partial t^{n}}\frac{\partial^{k_1}}{\partial u_1^{k_1}}\frac{\partial^{k_2}}{\partial u_2^{k_2}}\,\Phi(t_1,t_2,u)|_{t_1=t_2=u=0}.
\end{equation}
We note that the multivariate Bell polynomials of this type can be constructed from the standard (univariate) Bell polynomials,~\cite{Riordan:1946}
\begin{eqnarray}\label{eq:bell-construct}
    &&B_{n;k_1,k_2}(x_1^{(1)},x_1^{(2)},\dots;x_2^{(1)},x_2^{(2)},\dots)\\
    &&=\sum_{i=0}^n\binom{n}{i}\,B_{i,k_1}(x_1^{(1)},x_1^{(2)},\dots)\,B_{n-i,k2}(x_2^{(1)},x_2^{(2)},\dots).\nonumber
\end{eqnarray}

The first few Bell polynomials are
\begin{eqnarray}
    B_{1;1,0}(\{\vec{x}^{(i)}\})&=&x_1^{(1)},\nonumber\\
    B_{1;0,1}(\{\vec{x}^{(i)}\})&=&x_2^{(1)},\nonumber\\
    B_{2;1,0}(\{\vec{x}^{(i)}\})&=&x_1^{(2)},\nonumber\\
    B_{2;0,1}(\{\vec{x}^{(i)}\})&=&x_2^{(2)},\\
    B_{2;2,0}(\{\vec{x}^{(i)}\})&=&(x_1^{(1)})^2,\nonumber\\
    B_{2;0,2}(\{\vec{x}^{(i)}\})&=&(x_2^{(1)})^2,\nonumber\\
    B_{2;1,1}(\{\vec{x}^{(i)}\})&=&2\,x_1^{(1)}\,x_2^{(1)},\nonumber
\end{eqnarray}
and the corresponding Fa\`{a} di Bruno formula is~\cite{Riordan:1946}
\begin{eqnarray}\label{eq:faa-di-bruno-mult2}
    \frac{\partial^{n}}{\partial z^{n}} f(g_1(z),g_2(z))=\sum_{\substack{k_1,k_2\\ k_1+k_2\geq 0}}^{n}f^{(k_1,k_2)}\,B_{n;k_1,k_2}(\{g_1^{(i)};g_2^{(i)}\}).
\end{eqnarray}
Here the first few terms are given by
\begin{eqnarray}
    \frac{\partial}{\partial z}f(g_1(z),g_2(z))&=&f^{(1,0)}g_1^{(1)}+f^{(0,1)}g_2^{(1)}\,,\nonumber\\
    \frac{\partial^2}{\partial z^2}f(g_1(z),g_2(z))&=&f^{(1,0)}g_1^{(2)}+f^{(0,1)}g_2^{(2)}+f^{(2,0)}\big(g_1^{(1)}\big)^2\\
    &+&f^{(0,2)}\big(g_2^{(1)}\big)^2+2\,f^{(1,1)}\,g_1^{(1)}\,g_2^{(1)}\,.\nonumber
\end{eqnarray}

\section{Generalized Stirling numbers}\label{sec:app-b}

Consider multivariate Bell polynomials of the type (\ref{eq:gen-Bell-21}), with the derivatives $x^{(i,j)}$ given by $g^{(i,j)}$ in
(\ref{eq:gc-derivatives}). In this case, the generating function (\ref{eq:gen-function-multiv-Bell-21}) reduces to 
\begin{equation}
 \Phi(t_1,t_2,u)=\big[(1+\alpha_B\, t_1)\,(1+\alpha_{\bar B}\,t_2)\big]^{u/2},   
\end{equation}
and the corresponding Bell polynomials are
\begin{equation}
    B_{n,m;k}(\{g_3^{(i,j)}\})=\frac{(\alpha_B)^n\,(\alpha_{\bar B})^m}{2^k}\,\,s(n,m;k),
\end{equation}
where
\begin{equation}
    s(n,m;k)=\frac{1}{k!}\,\frac{\partial^n}{\partial z^n}\,\frac{\partial^m}{\partial w^m}\,\frac{\partial^k}{\partial u^k}\,\Psi(z,w,u)|_{z=w=u=0},
\end{equation}
are generalized Stirling numbers and 
\begin{equation}
    \Psi(z,w,u)=\big[(1+z)\,(1+w)\big]^u
\end{equation}
is the corresponding generating function.  

The generating function for the regular Stirling numbers of the first kind
\begin{equation}
    s(n,k)=\frac{1}{k!}\,\frac{\partial^n}{\partial z^n}\,\frac{\partial^k}{\partial u^k}\,\psi(z,w,u)|_{z=u=0}
\end{equation}
is of the form~\cite{Comtet:1974}
\begin{equation}
    \psi(z,u)=\big[1+z\big]^u.
\end{equation}
It follows that 
\begin{equation}\label{eq:gen-reg-stirling}
 s(n,0;k)=s(0,m;k)=s(n,k),   
\end{equation}
that $s(n,m;0)=0$, except for $n=m=0$, and that $s(n,m;k)=0$ for $k>n+m$ and that 
\begin{equation}
    s(n,m;k)=\sum_{l=0}^{k}\,s(n,k-l)\,s(m,l).
\end{equation}

Moreover, the generalized Stirling numbers satisfy the recurrence relations
\begin{eqnarray}\label{eq:gen-stirling-recurrence-1}
    s(n,m;k)&=&s(n-1,m;k-1)-(n-1)\,s(n-1,m;k),\\
    s(n,m;k)&=&s(n,m-1;k-1)-(m-1)\,s(n,m-1;k),\label{eq:gen-stirling-recurrence-2}
\end{eqnarray}
in close analogy to the one obeyed by the regular Stirling  numbers of the first kind~\cite{Comtet:1974}. The relation (\ref{eq:gen-stirling-recurrence-1}) holds for $n,k\geq 1$ and (\ref{eq:gen-stirling-recurrence-2}) for $m,k\geq 1$.
The two relations (\ref{eq:gen-stirling-recurrence-1}) and (\ref{eq:gen-stirling-recurrence-2}) combined yield a recurrence relation at fixed $k$,
\begin{equation}\label{eq:gen-stirling-recurrence-3}
    s(n,m;k)=s(n+1,m-1;k)+(n-m+1)\,s(n,m-1;k).
\end{equation}
The recurrence relation for factorial cumulants,
\begin{equation}
    F^{(n+1,m)}=\frac{\alpha_B}{\alpha_{\bar B}}\,F^{(n,m+1)}-(n-m)\,\alpha_B\,F^{(n,m)},
\end{equation}
which was empirically deduced in~\cite{Barej:2020ymr}, follows from (\ref{eq:gen-stirling-recurrence-3}).

The generalized Stirling numbers also have a combinatorial interpretation. The absolute value, $|s(n,m;k)|$, equals the number of permutations of $n$ blue beads and $m$ red beads in k disjoint cycles, where each cycle consists of only blue or only red beads. The phase of $s(n,m;k)$ reproduces the phase stemming from the derivatives of $g_3(x,\bar{x})$, given in (\ref{eq:gc-derivatives}).

\section{General relations between cumulants and factorial cumulants}\label{sect:App-c}

In this appendix we derive general relations between cumulants and factorial cumulants, which are independent of the assumed partition function. We start by considering the fluctuations of $N^{(tot)}=N_B+N_{\bar{B}}$. The cumulants $C_n$ and factorial cumulants $F_n$ are obtained by differentiating the corresponding generating functions,
\begin{equation}\label{eq:cumulant-gen-func-Ntot}
    C_n=\frac{d^n G(t)}{d\,t^n}\mid_{t=0}.
\end{equation}
and
\begin{equation}\label{eq:fac-cumulants-gen-func-Ntot}
    F_n=\frac{d^n H(x)}{d\,x^n}\mid_{x=1},
\end{equation}
which are closely related through~\cite{Kitazawa:2017ljq}
\begin{equation}\label{eq:cum-fac-cum-gen-func}
    G(t)=H(e^t),\qquad H(x)=G(\ln(x)).
\end{equation}
By applying the Fa\`{a} di Bruno formula to composite functions of the form $f(\ln x)$ and $f(e^t)$ one finds the following general relations between cumulants and factorial cumulants~\cite{Comtet:1974},
\begin{equation}\label{eq:Cn_to_Fn-relations}
    F_n=\sum_{k=1}^k\,s(n,k)\,C_k,\qquad C_n=\sum_{k=1}^n\,S(n,k)\,F_k,
\end{equation}
where $s(n,k)$ and $S(n,k)$ are Stirling numbers of the first and second kind, respectively. The Stirling numbers satisfy the orthogonality relations~\cite{Comtet:1974}
\begin{eqnarray}\label{eq:orth-Stirling-1}
    \sum_{k=0}^n\,S(n,k)\,s(k,m)&=&\delta_{n,m},\\
    \sum_{k=0}^n\,s(n,k)\,S(k,m)&=&\delta_{n,m}.\label{eq:orth-Stirling-2}
\end{eqnarray}

Analogously, one finds the relation between the cumulants and factorial cumulants of a subsystem by using the multivariant Fa\'{a} di Bruno formula (\ref{eq:faa-di-bruno-mult1}) for a function of the form $G(e^t,e^s)$,
\begin{equation}\label{eq:cumulant-fac-cumulant-relation-subsystem}
    C^{(n,m)}=\sum_{k_1=0}^n\,\sum_{k_2=0}^m\,F^{(k1,k2)}\,S(n,k_1)\,S(m,k_2).
\end{equation}
For completeness we note that the relation (\ref{eq:cumulant-fac-cumulant-relation-subsystem}) can be inverted, using the orthogonality property of the Stirling numbers, (\ref{eq:orth-Stirling-2}),
\begin{equation}\label{eq:fac-cumulant-cumulant-relation-subsystem}
    F^{(p,q)}=\sum_{n=0}^p\,\sum_{m=0}^q\,C^{(n,m)}\,s(p,n)\,s(q,m).
\end{equation}

Utilizing the fact that cumulants of the net baryon number correspond to cumulants  of  $\delta N_B - \delta N_{\bar B}$, one finds, using the binomial theorem, the general relation
\begin{equation}\label{eq:kappa-Cmn}
    \kappa_n=\sum_{i=0}^n\,\binom{n}{i}\,(-1)^{n-i}\,C^{(i,n-i)},
\end{equation}
where $\binom{n}{i}$ is a binomial coefficient. Similarly, one obtains the cumulants $\kappa^s_n$ of $\delta (N_B+N_{\bar B})$ in the subsystem using
\begin{equation}
    \kappa^s_n=\sum_{i=0}^n\,\binom{n}{i}\,C^{(i,n-i)}.
\end{equation}
In the limit $\alpha_B,\alpha_{\bar B} \to 1$, $\kappa_n\to 0$, while $\kappa^s_n \to C_n$. 

Finally, by inserting (\ref{eq:cumulant-fac-cumulant-relation-subsystem}) in (\ref{eq:kappa-Cmn}) one finds a general relation between the factorial cumulants and the net-baryon-number cumulants~\cite{Luo:2014rea},
\begin{equation}\label{eq:cumulant-factorial-cumulant-relation-appendix}
    \kappa_n=\sum_{\substack{k_1,k_2\\ k_1+k_2\geq 0}}^{n}F^{(k_1,k_2)}\,\sum_{i=0}^n\binom{n}{i}\,(-1)^{n-i}\,S(i,k_1)\,S(n-i,k_2).
\end{equation}

\end{document}